\documentstyle[11pt]{article}

\def\afour{
\setlength{\topmargin}{0mm}
\setlength{\headheight}{0mm}
\setlength{\headsep}{0mm}
\setlength{\textwidth}{6in}
\setlength{\textheight}{248mm}
\setlength{\oddsidemargin}{.25in}
\setlength{\evensidemargin}{.25in}
}

\newcommand\eq[1]{Eq.~(\ref{#1})}

\newcommand{\sub}[1]{_{\mbox{\scriptsize#1}}}

\newcommand\ee{\end{equation}}
\newcommand\be{\begin{equation}}
\newcommand\eea{\end{eqnarray}}
\newcommand\bea{\begin{eqnarray}}

\newcommand\TeV{\,\mbox{TeV}}

\newcommand\mone{^{-1}}

\newcommand\half{^{1/2}}

\newcommand\mpl{m_{Pl}}

\newcommand\lsim{\mathrel{\rlap{\lower4pt\hbox{\hskip1pt$\sim$}}
    \raise1pt\hbox{$<$}}}
\newcommand\gsim{\mathrel{\rlap{\lower4pt\hbox{\hskip1pt$\sim$}}
    \raise1pt\hbox{$>$}}}

\afour

\begin{document}

\begin{flushright}
LANCASTER-TH 94-14
hep-ph/9408324\\
\end{flushright}
\begin{center}
\Large
{\bf False Vacuum Chaotic Inflation: the New Paradigm?}

\vspace{.3in}
\normalsize
\large{David H. Lyth$^{\dagger}$ and Ewan D. Stewart$^*$} \\
\normalsize

\vspace{.6 cm}
{\em $^{\dagger}$School of Physics and Materials, \\ University of
Lancaster, \\ Lancaster LA1 4YB,~~~U.~K.}\\

\vspace{.4cm}
{\em  Department of Physics, Kyoto University, Kyoto 606,
Japan.}

\vspace{.6cm}
Contribution to appear in the proceedings of
`Birth of the Universe and Fundamental Physics'
held in Rome, 1994. Presented by D. H. Lyth.
\end{center}

\vspace{.6 cm}
Since the observation by COBE of cmb anisotropy at the expected
level, the idea that an initial inflationary
era sets the conditions for the subsequent hot big bang has enjoyed
unprecedented popularity. At the same time, the level of interest in
building an actual model of inflation has never been lower.
Perhaps the general feeling is that there are already
too many theoretically viable models, making it pointless
to look for more. Such a perception, if it exists, is far from the truth
because in the context of supergravity (now
generally accepted within the particle physics community)
the possibilities for model building are extremely limited, and largely
unexplored. Some recent work is reported here, with particular focus on
a remarkable new paradigm for model building [1-10],
in which one or more non-inflaton fields are sitting in
a false vacuum, whose energy density is supposed to dominate the total.

A model of inflation is defined by giving the
Lagrangian of the relevant fields, together with some information
about their initial values.
Here we focus on
Einstein gravity models, noting that modified gravity models can
usually be rewritten as Einstein gravity models at the expense
of making the matter lagrangian more complicated.
During inflation, the
energy density is
dominated by the scalar field effective potential, and
all fields except the inflaton field are constant.

There have been
three popular paradigms for Einstein gravity
inflation. `New Inflation' \cite{new}
postulates that the inflaton field $\phi$ is in
thermal equilibrium before inflation, at a temperature high enough to set
$\phi=0$. The minimum of the potential is supposed to correspond to
$\phi=\phi\sub{min}\neq0$, and inflation occurs as $\phi$ rolls slowly
towards $\phi\sub{min}$, ending when $\phi$ starts to oscillate around
$\phi\sub{min}$. After perhaps a long delay, reheating occurs
(ie., $\phi$ decays into other fields which thermalise).
The popularity of New Inflation has declined more or less continuously
since it was proposed,
because model building is difficult, and
because this paradigm leaves unresolved the question of what
sets the initial conditions for the pre-inflationary hot big bang.

`True Vacuum Chaotic Inflation' \cite{CHAOTIC}
postulates that our universe
emerges from the Planck scale with its
energy density dominated by scalar fields. Their values are large
enough that the potential soon dominates spacetime gradient terms.
 All of them but the inflaton field
quickly adjust to their true vacuum values and thereafter
the situation is the same as in New Inflation;
inflation occurs as the inflaton field $\phi$ rolls slowly
towards the vacuum and it ends when $\phi$ oscillates around the
vacuum, then after perhaps a long delay `reheating' occurs .
(Of course the
term `reheating' is a misnomer now, because thermal equilibrium is being
established for the first time).

`False Vacuum Chaotic Inflation' \cite{andrei,andrei1}
(also called `Hybrid Inflation') again
postulates initial domination by
scalar fields, with all except the inflaton field quickly
adjusting to minimise the potential. The difference is that
one or more of the non-inflaton fields is held in a false vacuum
through its coupling to the inflaton field.
The false vacuum energy dominates the total,\footnote
{A variant is to
have the false vacuum energy density negligible compared with the total.
As far as inflation is concerned one then has a scenario which is
indistinguishable from True Vacuum Chaotic Inflation,
but topological defects produced when the false vacuum destabilizes
might be cosmologically significant \cite{edet}.}
and inflation typically ends only
when the false vacuum is destabilized as the inflaton
field falls through some critical
value $\phi_c$. There ensues a phase transition to the
true vacuum, which may be of first or second order and which may or may
not produce topological defects.
(The first order case is discussed
by David Wands in these proceedings, and both cases are treated in some
detail in \cite{edet}.)

In global supersymmetry the (F-term) potential is of the form
\be
V=V\sub{global}\equiv\sum_\alpha|\partial W/\partial \phi_\alpha|^2
\label{global}
\ee
where the superpotential $W$ is an analytic function of the
complex scalar
fields $\phi_\alpha$.
Superpotentials have been constructed in the past which can lead to
New or True Vacuum Chaotic inflation \cite{olive}.
Recently, it was realised \cite{edet} that a superpotential
{\em already proposed in the literature in the context of particle
physics} leads naturally
to False Vacuum Chaotic Inflation. It is
\be
W=\sigma(\Psi_1\Psi_2+\Lambda^2/\sigma)\Phi
\ee
where $\Phi$, $\Psi_1$ and $\Psi_2$ are chiral superfields,
$\sigma\lsim1$ is a dimensionless coupling and $\Lambda$ sets the
energy scale.
The corresponding potential is
\be
V=\sigma^2|\psi_1\psi_2+\Lambda^2/\sigma|^2+\sigma^2
(|\psi_1|^2+|\psi_2|^2)|\phi|^2\,.
\ee
where $\phi$, $\psi_1$ and $\psi_2$ are the
scalar components of the superfields. For initial values
$|\phi|>|\psi_1|,|\psi_2|$
this gives False Vacuum Chaotic Inflation with
an absolutely flat potential $V(\phi)=\Lambda^4$.
A small slope can be provided \cite{edet} by giving $\phi$
a soft supersymmetry breaking mass $m\sim 1\TeV$, and in any case
a slope is provided by
the one loop correction which is
of order \cite{qaisar}
$(8\pi^2)\mone\sigma^2\Lambda^4
\ln[(8\pi/\mpl^2)\half\phi]$.

Globally supersymmetric models of inflation are
generally spoiled when supergravity is taken into
account \cite{edet,ewan1}. In supergravity the potential  \eq{global} becomes

\begin{eqnarray}
V &=& \exp \left( \frac{8\pi}{m_{{\rm Pl}}^2} K \right)
\left[
\sum_{\alpha,\beta} \left(
\frac{\partial^2 K}{\partial \bar{\phi}_{\alpha} \partial \phi_{\beta} }
	\right)^{-1} 	A_\alpha \bar A_\beta
- 3 \frac{8\pi}{m_{{\rm Pl}}^2} |W|^2 \right]
\label{sg1}\\
A_\alpha&\equiv&
\frac{\partial W}{\partial \phi_{\alpha} }
	+ \frac{8\pi}{m_{{\rm Pl}}^2} W
	\frac{\partial K}{\partial \phi_{\alpha} } \,.
\nonumber
\end{eqnarray}

Here the K\"{a}hler potential $K$ is a real
function of the scalar fields and their complex
conjugates.
The canonical form for K is
\begin{equation}
\label{sg2}
K = \sum_{\alpha} \left| \phi_{\alpha} \right|^2 + \ldots \,.
\end{equation}
which gives canonical kinetic terms to lowest order in the expansion
about $\phi=0$.
The corresponding expansion for the potential is
{\samepage
\begin{eqnarray}
V &=& \left( 1+\frac{8\pi}{m_{{\rm Pl}}^2} \sum_{\gamma}
	\left| \phi_{\gamma} \right|^2 + \ldots \right)
\left\{ \sum_{\alpha,\beta} \left( \delta_{\alpha \beta} + \ldots \right)
B_\alpha \bar B_\beta
- 3 \frac{8\pi}{m_{{\rm Pl}}^2} |W|^2 \right\} \nonumber\\
&=&
V_{\rm global} \left( 1 + \frac{8\pi}{m_{{\rm Pl}}^2} \sum_{\gamma}
	\left| \phi_{\gamma} \right|^2 + {\rm other\ terms} \right)
+ {\rm other\ terms}
\label{prob}\\
B_\alpha &\equiv&
\frac{\partial W}{\partial \phi_{\alpha} }
	+ \frac{8\pi}{m_{{\rm Pl}}^2}
	\left( \bar{\phi}_{\alpha} + \ldots \right) W \,.
\nonumber
\end{eqnarray}
}
This last expression shows that it
is difficult to build a supergravity model of inflation,
because it is difficult to satisfy the
necessary flatness condition \cite{physrep}
$|V''/V|   \ll 8\pi/\mpl^2$.
If the condition is satisfied in the global supersymmetric
limit,\footnote
{The opposite case where inflation does not occur in the global
susy limit is considered in \cite{ewan1}.}
it will be violated by the supergravity correction that we have
exhibited, for a generic inflaton field and
generic choices of $W$ and $K$.

How can this problem be avoided? One way is to suppose that the
inflaton corresponds to an `angular' as opposed to a `radial'
degree of freedom, so that $\sum_\alpha |\phi_\alpha|^2$ is fixed during
inflation. The only model so far proposed which achieves this is
`natural' inflation \cite{natural},
which invokes a sinusoidal inflaton potential
generated by instanton effects.

Barring this possibility, inflation can work only if the forms of
$W$ and $K$ are such that the
contribution to $V''/V$ of the exhibited term is cancelled.
Suitable forms have been written down for
New Inflation \cite{olive} and True Vacuum Chaotic Inflation
\cite{gonch,muram},
but they generally have no independent
motivation, and in particular do not emerge from superstrings.
Recently, forms for $W$ have been given that make
False Vacuum Chaotic
Inflation \cite{edet} and New Inflation \cite{kumem} work with
the minimal form for $K$ (no extra terms in \eq{sg2}),
\footnote{These both use the same cancellation, that occurs if the
effective superpotential during inflation is $W=\Lambda^2\phi$.}
but the minimal form also does not emerge from superstrings.
However, general conditions on $W$ and forms
for $K$ that allow inflation and
{\em do} emerge from superstrings
have recently been given \cite{edet,ewan1}.
The starting point of \cite{ewan1} is the following recipe for
ensuring that the potential receives no
inflaton-dependent supergravity correction.

Divide the fields $\phi_\alpha$ into two sets, $\varphi_i$ and
$\psi_n$, and suppose that there is an $R$ parity ensuring
that $W$ is an odd function of the $\psi_n$ and $K$ an even
function of the $\psi_n$.
Suppose that during inflation the $\psi_n$ are zero
(a natural value since the necessary condition $\partial
V/\partial \psi_n=0$ is then guaranteed by the $R$ symmetry).
Then the $R$ parity ensures that during inflation
$W=\partial W/\partial \varphi_i=0$.
Given these conditions, it is easy to show that there are no inflaton
dependent corrections to the global supersymmetry potential
provided that $K$ is of the form (suppressing subscripts)
\be
K=-\ln\left[f(\varphi,\overline\varphi)-\overline\psi
C(\chi,\overline\chi)\psi\right] + g(\chi,\overline\chi)
+{\cal O}\left(\psi^2,\overline\psi^2\right)
\,.
\ee
where the $\chi_j$ are a subset of the $\phi_i$ that are constant
during inflation ({\mbox i.e.} that do not contain the inflaton),
and $C_{mn}$ is a hermitian matrix.

\frenchspacing

\end{document}